# Tunable discrete scale invariance in transition-metal pentatelluride flakes

Yanzhao Liu[1], Huichao Wang[2], Haipeng Zhu[3], Yanan Li[1], Jun Ge[1], Junfeng Wang[3], Liang Li[3], Ji-Yan Dai[4], Jiaqiang Yan[5], David Mandrus[5,6], Robert Joynt[7,8] & Jian Wang[1,9,10*]

[1]International Center for Quantum Materials, School of Physics, Peking University, Beijing 100871, China.
[2]School of Physics, Sun Yat-Sen University, Guangzhou 510275, China.
[3]Wuhan National High Magnetic Field Center, Huazhong University of Science and Technology, Wuhan 430074, China.
[4]Department of Applied Physics, The Hong Kong Polytechnic University, Kowloon, Hong Kong, China.
[5]Materials Science and Technology Division, Oak Ridge National Laboratory, Oak Ridge, Tennessee 37831, USA.
[6]Department of Materials Science and Engineering, University of Tennessee, Knoxville, Tennessee 37996, USA.
[7]Kavli Institute of Theoretical Sciences, Chinese Academy of Sciences, Beijing 100049, China.
[8]Department of Physics, University of Wisconsin-Madison, 1150 Univ. Ave., Madison WI 53706, USA.
[9]CAS Center for Excellence in Topological Quantum Computation, University of Chinese Academy of Sciences, Beijing 100190, China.
[10]Beijing Academy of Quantum Information Sciences, Beijing 100193, China.

* Corresponding author: Jian Wang (jianwangphysics@pku.edu.cn)

**Abstract**

Log-periodic quantum oscillations discovered in transition-metal pentatelluride give a clear demonstration of discrete scale invariance (DSI) in solid-state materials. The peculiar phenomenon is convincingly interpreted as the presence of two-body quasi-bound states in a Coulomb potential. However, the modifications of the Coulomb interactions in many-body systems having a Dirac-like spectrum are not fully understood. Here, we report the observation of tunable log-periodic oscillations and DSI in $ZrTe_5$ and $HfTe_5$ flakes. By reducing the flakes thickness, the characteristic scale factor is tuned to a much smaller value due to the reduction of the vacuum polarization effect. The decreasing of the scale factor demonstrates the many-body effect on the DSI, which has rarely been discussed hitherto. Furthermore, the cut-offs of oscillations are quantitatively explained by considering the Thomas-Fermi screening effect. Our work clarifies the many-body effect on DSI and paves a way to tune the DSI in quantum materials.

## INTRODUCTION

One of the most important concepts in the area of phase transitions is scale invariance. A scale invariant system reproduces itself on different temporal and spatial scales. This is described by the relation $f(kx) = k^\lambda f(x)$, where $k$ is an arbitrary parameter, $\lambda$ is the scaling dimension, and $f$ is a physical field. Discrete scale invariance (DSI) is a weaker case of the scale invariance, where a system only obeys the scale invariance for specific choices of $k$[1,2]. With a fundamental scaling ratio $\lambda$ and characteristic log-periodicity, DSI arises in various contexts, such as earthquakes, financial crashes, turbulence and so on[1]. After being introduced to bound-state problems of quantum systems by Vitaly Efimov in 1970[3], the DSI had been observed only in cold atom systems for a long time[4-7]. It had not previously been observed in the solid state. The discovery of Dirac materials has changed that situation[8-14]. Especially in the topological transition-metal pentatelluride ZrTe$_5$, the quantum oscillations with log-periodicity have revealed the existence of DSI in a solid state system[10]. The origin is attributed to the quasi-bound states induced by the supercritical Coulomb interaction between the massless Dirac fermions and charged impurities[10,12,13]. Further studies reported the log-periodic oscillations in HfTe$_5$[14] and elemental semiconductor tellurium[15], confirming that the DSI feature can be a universal characteristic of Dirac materials with Coulomb impurities.

ZrTe$_5$ and HfTe$_5$ are predicted to be quantum spin Hall insulators in the two-dimensional (2D) limit and the 3D crystals are located near the phase boundary between weak and strong topological insulators[16]. Later studies indicated that the topological natures of both ZrTe$_5$ and HfTe$_5$ are very sensitive to the crystal lattice constant and detailed composition[17,18]. Taking ZrTe$_5$ as an example, some angle-resolved photoemission spectroscopy (ARPES)[19,20] and magneto-infrared spectroscopy studies show that ZrTe$_5$ is a Dirac semimetal[21]. The observed negative magnetoresistance (MR) that is related to the chiral anomaly and the anomalous Hall effect support the hypothesis of a massless Dirac band structure[22-24]. However, other ARPES and scanning tunneling microscopy (STM) results suggest that ZrTe$_5$ is a topological insulator[25-28]. Thus, the transition-metal pentatelluride ZrTe$_5$ and HfTe$_5$ are ideal platforms to investigate different intriguing physical properties due to their high tunability[29-34]. In particular, the very small Fermi surface of the compounds has enabled some peculiar findings in the ultraquantum regime, such as log-periodic quantum oscillations and three-dimensional quantum Hall effect[10, 14, 35]. As described above, the log-periodic oscillations revealing the DSI feature can be convincingly explained by a two-body quasi-bound state model. However, the solids ZrTe$_5$ and HfTe$_5$ are in fact many-body systems, and the Coulomb interactions are modified by screening effects that are not fully understood. The screening of the Coulomb interaction is closely related to the carrier density. In the experiments we find that the carrier density in ZrTe$_5$ flakes changes with thickness[29]. The interaction between the layers of ZrTe$_5$ and HfTe$_5$ is comparable to graphene, making it easy to get flakes from the bulk samples by exfoliation[16]. Thus, it is interesting to study the log-periodic quantum oscillations in the transition-metal pentatelluride flakes with different carrier density by thickness control, which may provide insights into many-body effects on DSI.

In this work, we carried out systematic magnetotransport measurements on ZrTe$_5$ and HfTe$_5$ flakes under magnetic fields ($B$) up to 53 Tesla (T). The log$B$-periodic oscillations can be detected in both ZrTe$_5$ and HfTe$_5$ flakes with thicknesses down to about 160 nm. In the thick ZrTe$_5$ flake (~ 1200 nm), log$B$ oscillations with a scale factor $\lambda$ ~ 3.3 is observed. The consistent scale factors between the flake ($\lambda \sim 3.3$) and bulk crystals ($\lambda \sim 3.2$)[10] show that the quality of the samples is not affected by the process of exfoliation and flake fabrication. More interestingly, a relatively smaller scale factor $\lambda \sim 1.6$ can be extracted in the MR and Hall traces of flakes with a thickness of about 160 nm. Theoretically, this index $\lambda$ decreases with increasing the effective charge of impurities that compose quasi-bound states responsible for the DSI and the log-periodicity. Further analysis shows that the carrier density in the flakes increases as the thickness decreases. We believe that the high carrier density makes the vacuum polarization effect weaker and leads to a larger effective charge, which results in a smaller scale factor. Thus, this work reveals a thickness-tuned scale factor of log$B$ quantum oscillations in ZrTe$_5$ and HfTe$_5$ flakes and provides a perspective on the DSI in solid-state systems.

**RESULTS**

**Temperature dependence of resistance**

ZrTe$_5$ and HfTe$_5$ belong to the orthorhombic space group *Cmcm* ($D_{2h}^{17}$)[36]. Figure 1(a) shows the crystal structure of ZrTe$_5$ and HfTe$_5$. Within the *a-c* plane, the trigonal prismatic chains of "ZrTe$_3$" or "HfTe$_3$" run along the *a* axis and are linked by parallel zigzag chains of "Te$_2$" along the *c* axis. The layers of ZrTe$_5$ and HfTe$_5$ are stacked along the *b* axis. The false-color scanning electron microscopy (SEM) image of a fabricated ZrTe$_5$ flake (grey color) with six electrodes (gold color) is displayed in Fig. 1(b). A schematic measurement structure of standard six-electrode-method is illustrated in the inset of Fig. 1(c). The current is applied along the *a* axis and the magnetic field is along the *b* axis for all measurements.

Figure 1(c) represents typical resistance-temperature (*R-T*) curves of HfTe$_5$ (~300 nm thick)

and ZrTe$_5$ (~190 nm thick) flakes. Resistance peaks at $T_p$ ~ 65 K and 74 K can be detected for HfTe$_5$ and ZrTe$_5$ flakes, respectively. It is noted that the $T_p$ of the flakes are definitely higher than that of the bulk[10, 14]. The origin of the resistance peak in ZrTe$_5$ and HfTe$_5$ has been discussed for decades, and recently the view of the Lifshitz transition during the changing of temperatures explains some findings in the compounds[28]. However, in other reports, the Hall resistances of ZrTe$_5$ and HfTe$_5$ remain $p$ type with increasing temperatures, which is not consistent with the picture of a Lifshitz transition[29, 37]. An alternative two-band model has been presented to be responsible for the transport anomaly and the various $T_p$ in different samples[14, 29]. In this picture, at low temperatures, the metallic $R$-$T$ curve mainly relies on a semi-metallic Dirac band, while a semiconducting band dominates at higher temperatures. The combination and competition between the two bands lead to the resistance peak at $T_p$. In exfoliated flakes, the carrier density of the Dirac holes commonly increases with decreasing thickness, indicating a shift of the Fermi level towards the valence band. Thus, the thickness decrease can alter the competition balance and result in a higher $T_p$.

**Tunable log-periodic oscillations and DSI**

A ZrTe$_5$ flake labeled s1 with a thickness of about 1200 nm was measured under magnetic fields up to 53 T and the MR curves in a semilogarithmic scale are shown in Fig. 2(a). By subtracting a smooth background, the log$B$-periodic oscillating components can be extracted and are shown in Fig. 2(b). For clarity, data curves in Figs. 2(a) and 2(b) are shifted. Distinct log-periodic magneto-oscillations can be observed at low temperatures. With increasing temperature, thermal broadening eventually becomes comparable to the intrinsic width of the quasi-bound states that induce log-periodic oscillations[10, 14]. Thus the oscillations gradually attenuate with increasing temperature and finally disappear above 120 K (see Supplementary Figure 1(a)-(b)), which agree with the observations in bulk crystals[10]. Consistent results were also acquired by using the second differential method to extract the oscillations, as shown in Supplementary Figure 1(d). The independence of the oscillations on the subtraction methods demonstrates that the log-periodic structures are intrinsic properties of the materials (See Supplementary Note 1 and Supplementary Figures 1-5 for more details). The characteristic magnetic fields ($B_n$) of the observed oscillation peaks and dips can be indexed as $n$ and ($n$-0.5), respectively. By plotting log ($B_n$) as a function of $n$, a linear dependence is revealed to confirm the log-periodic property (Fig. 2(c)). Based on the linear fitting in Fig. 2(c), a dominant scale factor $\lambda = B_n/B_{n+1} \sim 3.34$ can be obtained. Further, the Fast Fourier Transform (FFT) result of the log-periodic oscillations is shown in Fig. 2(d). The sharp FFT frequency peak is located at $F \sim 1.87$, which indicates a period of log ($B_n$) ~ 0.53 and a main scale factor $\lambda \sim 3.38$. The FFT shows some broadening, with a full width at half maximum (FWHM) analysis giving a reasonable $\lambda$ range of about [2.58, 5.98]. It is noted that the properties of the oscillations in the 1200 nm thick $ZrTe_5$ flake are very close to those observed in the bulk[10], showing that the quality of the device is still reserved during the process of the thin flake fabrication.

The thinner flakes are further studied for comparison. Figure 3 shows the log-periodic oscillations in the MR and Hall traces of the $ZrTe_5$ flake (s2) with a thickness of about 137 nm. The oscillatory part of MR at 4.2 K after subtracting a smooth background is displayed in Fig. 3(a). The inset of Fig. 3(a) plots the MR of s2 at 4.2 K as a function of the magnetic field. The log-periodic oscillations can be observed and survive up to 120 K (Supplementary Figure 6).

The index plot and FFT result of the oscillations are shown in Fig. 3(b). Both the linear fitting and the sharp FFT frequency peak confirm the log-periodicity and give a $\lambda$ ~1.60. The FWHM of the frequency peak further indicates a range for $\lambda$ of [1.45, 3.61]. The log-periodic oscillations are also observed in the Hall traces of the s2, as shown in Figs. 3(c) and 3(d). We subtract the background of the Hall signal by using similar methods to the MR. $\Delta R_{yx}$ as a function of log $B$ is plotted in Fig. 3(c). Figure 3(d) displays the linear fitting of the index and the FFT analysis of the data shown in Fig. 3(c). A scale factor $\lambda$ ~1.66 can be obtained for the oscillations in the Hall resistance, coinciding with $\lambda$~1.60 for the MR. The near equality of $\lambda$ in the MR and the Hall resistance is in agreement with the theoretical expectation[14]. Further measurements on a $ZrTe_5$ flake with a thickness of about 50 nm show log-periodic quantum oscillations in Hall trace with $\lambda$~1.52 (Supplementary Figure 7). The decrease of the scale factor is also supported by the observations in the $HfTe_5$ flake, which shows clearer oscillations. Figure 4 displays the magnetotransport results of the $HfTe_5$ flake (s3) with a thickness of about 180 nm. By using the same background subtraction method, the log-periodic oscillations in MR and Hall traces can be extracted, as shown in Figs. 4(a) and 4(c). The oscillations in MR at high magnetic fields can survive up to 78 K (Supplementary Figure 8). Figs. 4(b) and 4(d) plot the linear fitting and FFT analysis for the oscillations in the MR and the Hall trace, respectively. The scale factor $\lambda$ in s3 is estimated to be around 1.6, smaller than $\lambda$~3.5 in the bulk[14].

**DISCUSSION**

As discussed in previous works, the log-periodic quantum oscillations can be attributed to the two-body quasi-bound states composed of a massless Dirac particle and a stationary attractive

Coulomb impurity[10, 14]. For DSI to occur, we need the supercritical condition $\alpha > 1$ to occur in the system, where $\alpha = Ze^2/\varepsilon\hbar v_F$ is the effective fine-structure constant of the impurity. Here $Z$ is the ionicity, $\varepsilon$ is the background dielectric constant and $v_F$ is the Fermi velocity. When the supercritical condition holds, the radius of the quasi-bound states of the impurity is discrete scale invariant according to the relation $\frac{R_{n+1}}{R_n} = \sqrt{\lambda} = e^{\frac{\pi}{s_0}}$ with $s_0 = \sqrt{\alpha^2 - 1}$. $R_n$ is the characteristic radius of the quasi-bound state and when the magnetic length $\ell_n = \sqrt{\frac{\hbar c}{eB_n}} = s_0 R_n$ under an applied magnetic field, the energy of the $n$-th quasi-bound states rises rapidly and moves through the Fermi energy. Hence the sequence of fields at which there is a peak in the density of states at the Fermi energy satisfies $\frac{B_{n+1}}{B_n} = e^{2\pi/s_0} = e^{2\pi/\sqrt{\alpha^2-1}} = \lambda$[10]. When $B = B_n$, the resonant scattering between the mobile carriers and the quasi-bound states influences the transport properties and results in a log-periodic correction to the MR and Hall traces.

The most interesting feature of our results is that compared to the bulk or thick flakes of $ZrTe_5$ and $HfTe_5$, the scale factor $\lambda$ in the thinner flakes becomes quite small. As mentioned above, $\lambda$ is a monotonically decreasing function of $\alpha$ and $\alpha = Ze^2/\varepsilon\hbar v_F$. The Fermi velocity $v_F$ is expected to be the same in bulk and thin flakes of a Dirac material. The background dielectric constant $\varepsilon$ depends on the core electrons and so it is also unlikely to change. Thus, we must look for changes in $Z$ to explain the change in $\lambda$.

Based on Figs. 2-4, we notice that the ratio of oscillation amplitude to MR, i.e., $\Delta R_{xx}(B)/R_{xx}(B)$, for the $n = 1$ peak in thick and thin flakes are 1.3% and 0.5%, respectively. The larger surface-volume ratio but weaker oscillations in thinner flakes may exclude the possibility that the log-periodicity arises from a surface effect. By estimating from the Hall traces (Supplementary Figure 1 and Figure 6), the Dirac hole carrier density of the 1200 nm thick ZrTe$_5$ flake is $2.6\times10^{15}$ cm$^{-3}$, very close to that of bulk ZrTe$_5$[10], whereas the Dirac carrier density of the 137 nm thick ZrTe$_5$ flake is $9.9\times10^{16}$ cm$^{-3}$. The greatly increased carrier density in the thinner flakes can have a profound effect on screening of the Coulomb interaction. There are three types of screening in the transition-metal pentatelluride compounds, including the usual effects arising from bound electrons and conduction carriers, as well as vacuum polarization. The screening from the bound core electrons is responsible for background dielectric constant $\varepsilon$. As mentioned above, this screening effect is unlikely to change between the bulk and flakes since it depends only on very local features of the atomic structure. The screening from conduction carriers occurs in ordinary metals and in Weyl metals. It is noticed that the screening effect does not change the effective charge (ionicity $Z$) and has little influence on $\lambda$. The vacuum polarization is a specific feature of Dirac and Weyl semimetals and comes from the fact that the joint density of states for particle-hole excitations is large at low energies[38, 39]. The introduction of an impurity charge excites many virtual particle-hole pairs in Dirac and Weyl semimetals, much more than in a typical insulator with a hard gap. In Ref.38 it was found in two dimensions that this effect renormalizes $Z$ downwards: $Z \rightarrow Z^* = Z / [1+ZQ \ln (1/\kappa a)]$, where $\kappa$ is proportional to the Fermi wavevector and thus to the square root of the carrier concentration $n_e$. $a$ is a short-distance cutoff at the atomic scale and $Q$ is a numerical factor of order 1. A review of interaction effects in the Coulomb impurity problem in graphene concluded that this is the main effect[40]. There is no such complete theory in three dimensions to our knowledge, but the physics is expected to be similar. The renormalization of $Z$ is substantial due to its large starting value and indeed if $n_e a^2$ is of order $10^{-4}$, as in our thin flakes, then changes in $\lambda$ by a factor of 2 are easily obtained. Essentially, in the thinner flakes the vacuum polarization effect decreases and the effective charge becomes larger, which gives an increase in $s_0$ and a smaller $\lambda$. Therefore, we propose that vacuum polarization plays a crucial role in the decrease of $\lambda$ in flakes. Further theoretical efforts are still needed to make this connection fully quantitative.

Although at 4.2 K there are more than three oscillating cycles detected in the thin flakes due to the smaller scale factor, the oscillation signal becomes clear only above 3-5 T for the thin flakes, much higher than that for thick flakes or bulk[10]. We explain that the critical field $B_c$ over which the oscillations should be observed is closely related to the Thomas-Fermi screening length $\xi$. Though the screening from conduction carriers has little influence on $\lambda$, it can affect the visibility of the oscillations by converting the Coulomb potential $Ze^2/r$ to $\frac{Ze^2}{r} \times e^{-r/\xi}$. When the distance from the impurity is greater than $\xi$, the Coulomb potential becomes exponential and DSI is lost. In the thinner flakes, the increased carrier density leads to decreased $\xi$ accompanying with a larger $B_c$. For the thinner flakes, the calculated $B_c \sim 3.5$ T and the number of oscillations $N \sim 5$ are consistent with the experimental results (see Supplementary Note 2 for more details).

In summary, we carried out systematic transport measurements on the transition-metal pentatelluride $ZrTe_5$ and $HfTe_5$ flakes with different thicknesses. Log$B$-periodic oscillations and DSI feature are detected in both the MR and the Hall traces of these flakes. The scale factor is observed to decrease with decreasing thickness, which is attributed to the weaker vacuum polarization effect resulting from higher carrier density. This work offers a way to tune the DSI and log-periodic oscillations in quantum materials. It also provides deep insights into many-body effects with a Dirac-like spectrum.

## METHODS

### Sample information

The high-quality ZrTe$_5$ and HfTe$_5$ single crystals were grown by the Te-flux method as described in previous reports[10,14,37]. The crystals were characterized by powder X-ray diffraction, scanning electron microscopy with energy dispersive X-ray spectroscopy, and transmission electron microscopy.

### Device fabrication

The ZrTe$_5$ and HfTe$_5$ flakes were exfoliated by using the Scotch tape method onto 300 nm-thick SiO$_2$/Si substrates. After spin coating of poly (methyl methacrylate) (PMMA), the standard electron beam lithography in a FEI Helios NanoLab 600i Dual Beam System was carried out to define electrodes. Metal electrodes (Pd/Au, 6.5/300 nm) were deposited in a LJUHV E-400L E-Beam Evaporator after Ar plasma cleaning.

### Transport measurements

Electronic transport measurements in this work were mainly conducted in the pulsed high magnetic field facility (53 T) at Wuhan National High Magnetic Field Center. Standard six-electrode-method was used for the measurements. The current is applied along the *a* axis and the magnetic field is along the *b* axis for all measurements. Gold wires are attached to the electrodes by the silver epoxy and fixed on the substrate by GE varnish to avoid the vibration under pulsed field (see Supplementary Figure 9).

## DATA AVAILABILITY

The data that support the findings of this study are available from the corresponding author on reasonable request.

**CODE AVAILABILITY**

The code in this work is available from the corresponding author on reasonable request.

**ACKNOWLEDGMENT**

We thank Liqin Huang, Zihan Yan, Qingzheng Qiu, Haiwen Liu, Yongjie Liu for discussions on the data. This work was financially supported by the National Key Research and Development Program of China (2018YFA0305604 and 2017YFA0303302), the National Natural Science Foundation of China (Grant No. 11888101, Grant No. 11774008, Grant No. 12004441), Beijing Natural Science Foundation (Z180010) and the Strategic Priority Research Program of Chinese Academy of Sciences (Grant No. XDB28000000). J.Y. and D.M. were supported by the U.S. Department of Energy, Office of Science, Basic Energy Sciences, Materials Sciences and Engineering Division. H.W. acknowledges the support of the Hundreds of Talents program of Sun Yat-Sen University and the Fundamental Research Funds for the Central Universities (No. 20lgpy165).

**COMPETING INTERESTS**

The authors declare no competing interests.

**AUTHOR CONTRIBUTIONS**

Jian Wang conceived and supervised the research. Y. Liu, H. W. performed the transport measurements. Y. Li and J. G. fabricated devices. J. Y. and D. M. grew the $ZrTe_5$ and $HfTe_5$ bulk crystals. H. Z., Junfeng Wang, L. L., and J.-Y. D. helped in the transport measurements. R. J. carried out theoretical calculations. Y. Liu, H. W. and Jian Wang analyzed the data. Y. Liu, H. W., R. J. and Jian Wang wrote the manuscript with input from all authors.

These authors contributed equally: Yanzhao Liu, Huichao Wang.

## ADDITIONAL INFORMATION

**Supplementary information** is available for this paper at https://doi.org/xxx.

**Correspondence** and requests for materials should be addressed to Jian Wang.

**FIGURE LEGENDS**

**Fig. 1 Crystal structure and resistance-temperature characteristics of the transition-metal pentatelluride flakes.** (**a**) Crystal structures of $ZrTe_5$ and $HfTe_5$. (**b**) Scanning electron microscope false colour image of a typical $ZrTe_5$ flake. Scale bar represents 10 μm. (**c**) Typical temperature-dependent longitudinal resistance of the transition-metal pentatelluride flakes. The peak anomaly appears at $T_p \sim$ 65 K and 74 K for $HfTe_5$ (300 nm thick) and $ZrTe_5$ (190 nm thick) flakes, respectively. Inset shows a schematic structure for the electrical transport measurements.

**Fig. 2 Log-periodic MR oscillations and DSI in the $ZrTe_5$ flake (s1) with a thickness of about 1200 nm.** (**a**) MR of s1 at 4.2 K and 15 K versus magnetic field. (**b**) Extracted MR oscillatory part from the data in (**a**). (**c**) Log$B$-periodicity of the MR oscillations in s1. (**d**) FFT result of the MR oscillations at 4.2 K in (**b**).

**Fig. 3 Log-periodic oscillations and DSI in the $ZrTe_5$ flake (s2) with a thickness of about 137 nm.** (**a**) Extracted oscillatory part from MR of s2 at 4.2 K. Inset: MR of s2 at 4.2 K. (**b**) Log$B$-periodicity of the MR oscillations in s2. Inset: FFT result of the MR oscillations at 4.2 K in (**a**). The scale factor $\lambda$ of the oscillations in the MR is about 1.60. (**c**)-(**d**) are results for the Hall trace of s2 at 4.2 K. The scale factor $\lambda$ of the oscillations in the Hall trace is about 1.66, coincided with that for the MR.

**Fig. 4 Log-periodic oscillations and DSI in the $HfTe_5$ flake (s3) with a thickness of about**

**180 nm.** (**a**) Extracted oscillatory part from MR of s3 at 4.2 K. Inset: MR of s3 at 4.2 K. (**b**) Log$B$-periodicity of the MR oscillations in s3. Inset: FFT result of the MR oscillations at 4.2 K in (**a**). (**c**)-(**d**) are results for the Hall traces of s3 at 4.2 K. The scale factor $\lambda$ for the log-periodic oscillations in the Hall trace is about 1.59, consistent with that (~1.62) for the MR.

Figures
Figure 1

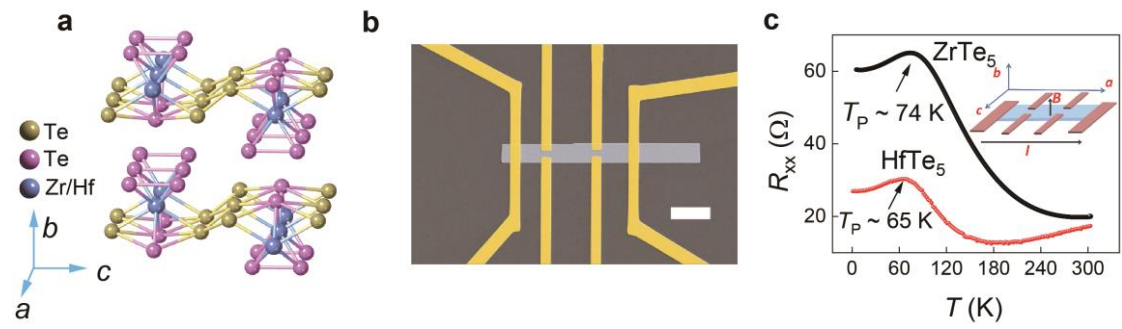

Figure 2

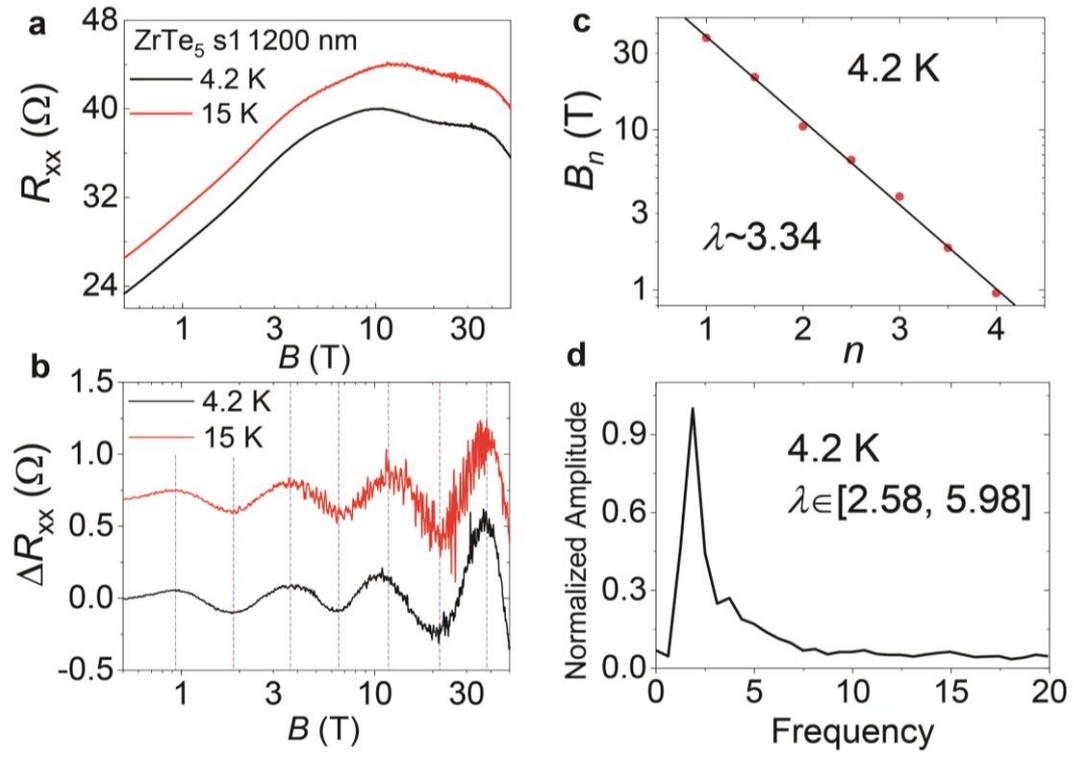

Figure 3

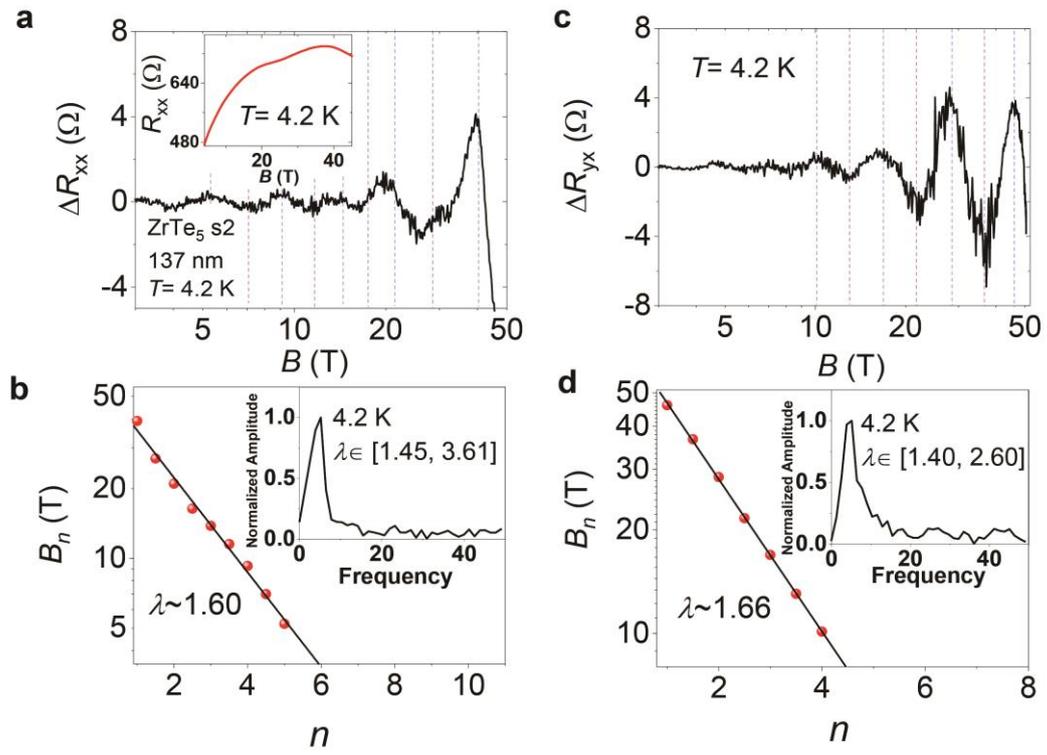

Figure 4

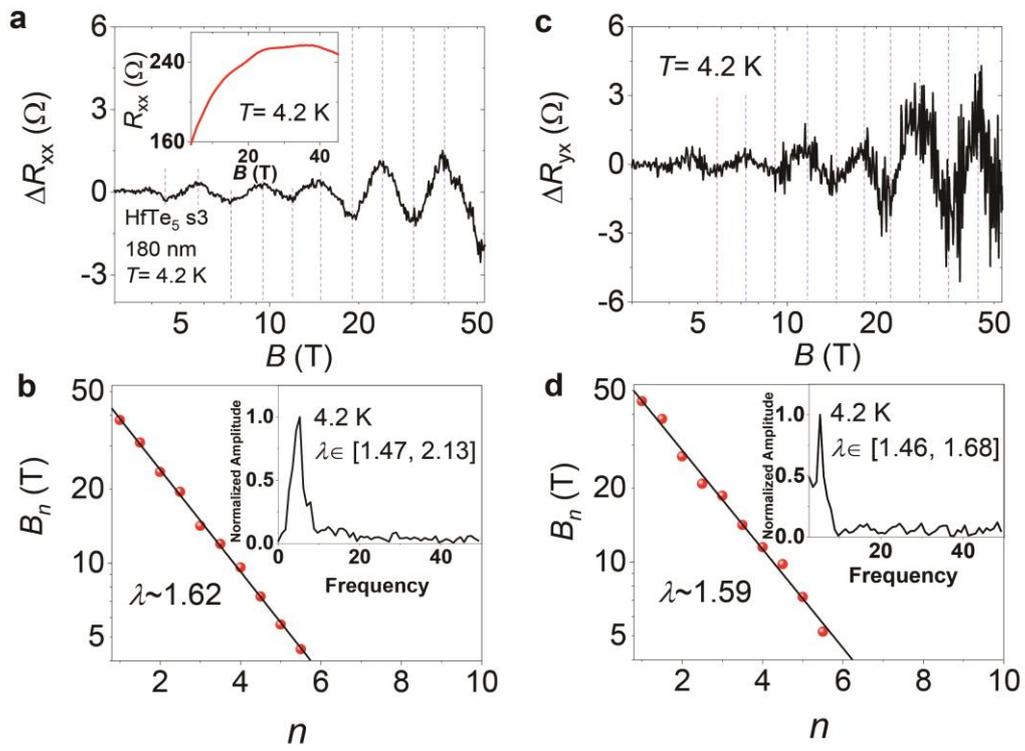

Supplementary Information for

# Tunable discrete scale invariance in transition-metal pentatelluride flakes

Yanzhao Liu[1], Huichao Wang[2], Haipeng Zhu[3], Yanan Li[1], Jun Ge[1], Junfeng Wang[3], Liang Li[3], Ji-Yan Dai[4], Jiaqiang Yan[5], David Mandrus[5,6], Robert Joynt[7,8] & Jian Wang[1,9,10*]

[1]International Center for Quantum Materials, School of Physics, Peking University, Beijing 100871, China.
[2]School of Physics, Sun Yat-Sen University, Guangzhou 510275, China.
[3]Wuhan National High Magnetic Field Center, Huazhong University of Science and Technology, Wuhan 430074, China.
[4]Department of Applied Physics, The Hong Kong Polytechnic University, Kowloon, Hong Kong, China.
[5]Materials Science and Technology Division, Oak Ridge National Laboratory, Oak Ridge, Tennessee 37831, USA.
[6]Department of Materials Science and Engineering, University of Tennessee, Knoxville, Tennessee 37996, USA.
[7]Kavli Institute of Theoretical Sciences, Chinese Academy of Sciences, Beijing 100049, China.
[8]Department of Physics, University of Wisconsin-Madison, 1150 Univ. Ave., Madison WI 53706, USA.
[9]CAS Center for Excellence in Topological Quantum Computation, University of Chinese Academy of Sciences, Beijing 100190, China.
[10]Beijing Academy of Quantum Information Sciences, Beijing 100193, China.

* Corresponding author: Jian Wang (jianwangphysics@pku.edu.cn)

**Supplementary note 1.**

**Analysis of log-periodic quantum oscillations**

The log-periodic oscillations are superimposed in a large smooth magnetoresistance background. The oscillating signals are relatively weak, and thus we need to subtract the background as generally do for quantum oscillations[1]. We used two different methods to extract the oscillating term. As shown in Supplementary Figure 4, the oscillations can be obtained by subtracting a smooth background[2,3]. Alternatively, we computed the second derivative for the measured raw data to get the oscillations, which are shown in Supplementary Figures 1-3. The dip/peak positions of the oscillations are consistent in two methods. The independence of the oscillations on the subtraction methods demonstrates that the log-periodic structures are intrinsic properties in the materials. In addition, we measured a 5-nm-thick $Bi_2Se_3$ film in the same pulsed high magnetic field facility, as presented in Supplementary Figure 5. There is no oscillating feature in the entire magnetic field region up to 50 T. And the log-periodic oscillations are absent in some other materials measured in this facility[4,5]. These results rule out the possible origin of the oscillations from the measurement facilities. Based on the above analyses, we confirm that the log-periodic quantum oscillations are intrinsic properties of our flake samples.

**Supplementary note 2.**

**Number of Oscillations**

It is noticed that only a finite number of oscillations are observed in the experiments, even though the log-periodicity is accurately preserved. In this section, we show how to compute the field range over which the oscillations should be observed, and $N$, the total number of oscillations.

The basic idea is that there is a single oscillation for each bound state (labeled by *n*) when the magnetic length $\ell_n$ and the quasi-bound state radius $R_n$ coincide up to a factor:

$$\ell_n = \sqrt{\frac{\hbar c}{eB_n}} = s_0 R_n, \qquad (1)$$

with $s_0 = \sqrt{\alpha^2 - 1}$. Hence the number of oscillations is in principle the same as the number of bound states that satisfy DSI. This assumes that there are no limits on the field strength, which is of course not actually the case. We deal with this issue below.

There are natural cutoffs for $R_n$ at both ends. At the short-distance end there is $r_0$, the cutoff length close to the atomic scale. This is unfortunately not very accurately known. At the long-distance end we have the Thomas-Fermi screening length $\xi$. When the distance from the impurity is greater than $\xi$, the potential becomes exponential and DSI is lost. So the oscillations occur when the inequality $r_0 < R_n < \xi$ is satisfied. Inverting the equation above, we find that the oscillations occur over a field range

$$\frac{\hbar c}{s_0^2 e \xi^2} < B < \frac{\hbar c}{s_0^2 e r_0^2}. \qquad (2)$$

However, if $r_0 = 0.2$ nm, then the upper limit on $B$ is of order $10^3$ T, so it's always true in practice that the upper limit is set by the maximum experimental field $B_m$. In the current experiments, this is $B_m = 53$ T. To get the lower limit we need the screening length $\xi$, which is determined from the equations $\xi^{-2} = \frac{4\pi e^2 dn}{d\mu}, \mu = \hbar v_F k_F,$ and $n = g k_F^3/6\pi^2$. Here $n$ is the carrier density of holes from Dirac band, $\mu$ is the chemical potential and $g = 4$ is the number of Weyl nodes. Inserting the measured electron densities we find $\xi_b = 11.1$ nm for the thick (bulk-like) films and $\xi_f = 2.20$ nm for the thin films. Using the above formula, we find that the oscillations will be present for $B > B_{cb} = 0.79$ T for the thick film, while in thin films they only appear once $B > B_{cf} = 3.5$ T. Both of these values are in agreement with experiment.

The total number of oscillations is given for the thick films by

$$N_b = \frac{\log\left(\frac{B_m}{B_{cb}}\right)}{\log \lambda_b} = 3.5. \tag{3}$$

And for thin films by

$$N_f = \frac{\log\left(\frac{B_m}{B_{cf}}\right)}{\log \lambda_f} = 5.3. \tag{4}$$

In both cases, the values are consistent with experimental observations.

The data show convincingly that the DSI is cut off at long distances by the Thomas-Fermi screening length.

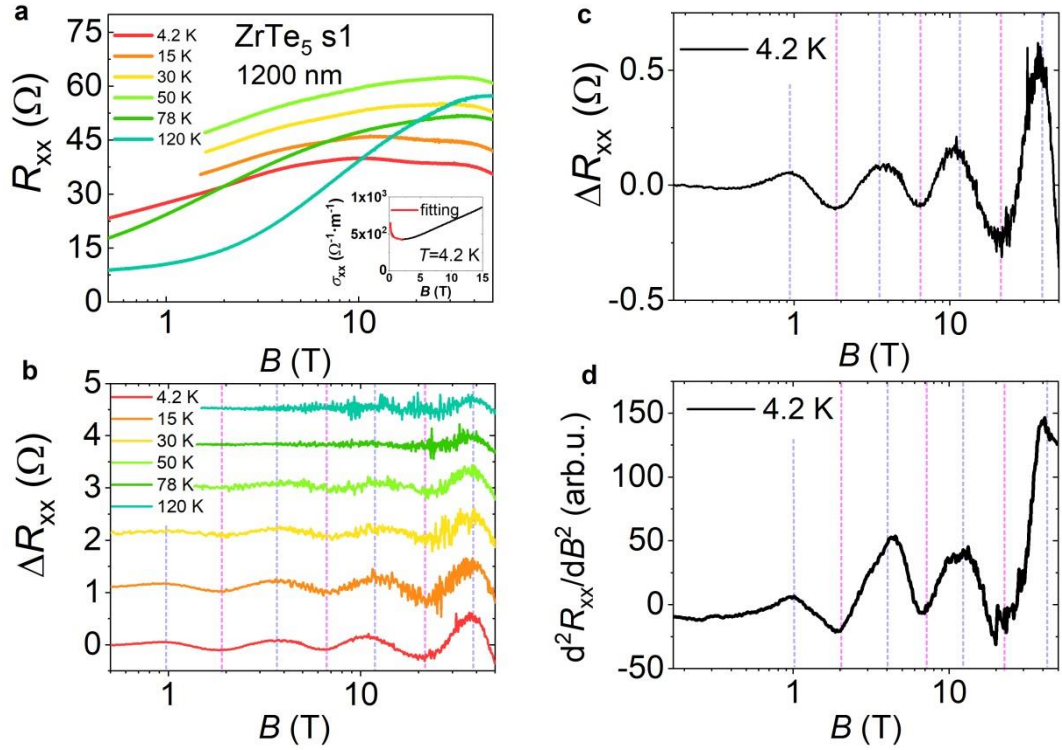

Supplementary Figure 1. Temperature-dependent log$B$-periodic oscillations in ZrTe$_5$ flake (s1) with a thickness of about 1200 nm. (a) MR of s1 at different temperatures. Inset: Two-carrier model fitting of s1 at 4.2 K. Red line is the fitting result. The Dirac hole density (mobility) is estimated to be $2.6\times10^{15}$ cm$^{-3}$ ($3.9\times10^5$ cm$^2$V$^{-1}$s$^{-1}$). (b) Log$B$-periodic oscillations in s1 at different temperatures. Data curves in (a) and (b) are shifted for clarity. (c) and (d) Extracted log$B$-periodic oscillations at 4.2 K by subtracting a smooth background (c) and the second derivative method (d). Dashed lines serve as guides to the eye.

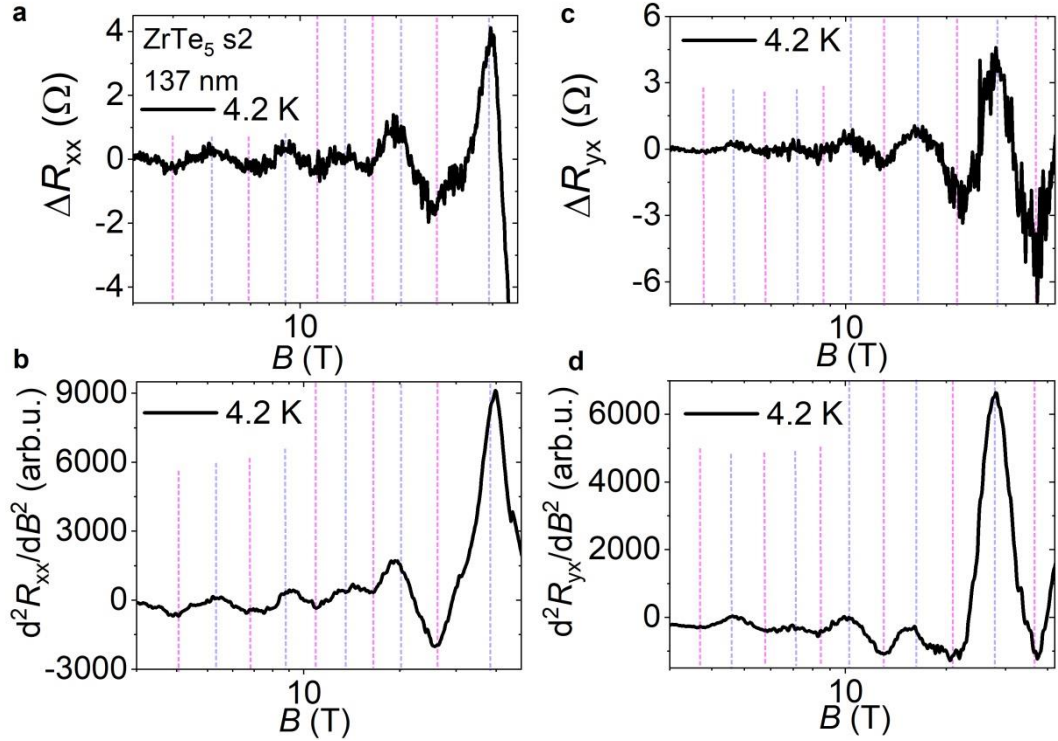

Supplementary Figure 2. Extracted log$B$-periodic oscillations in ZrTe$_5$ flake (s2) at 4.2 K. (a) Log$B$-periodic oscillations in MR of s2 at 4.2 K from the raw data after subtracting a smooth background. (b) The second derivative results of the raw MR data of s2 at 4.2 K. (c) Log$B$-periodic oscillations in Hall traces of s2 at 4.2 K after subtracting a smooth background. (d) The second derivative results of the raw Hall trace of s2 at 4.2 K. Dashed lines serve as guides to the eye.

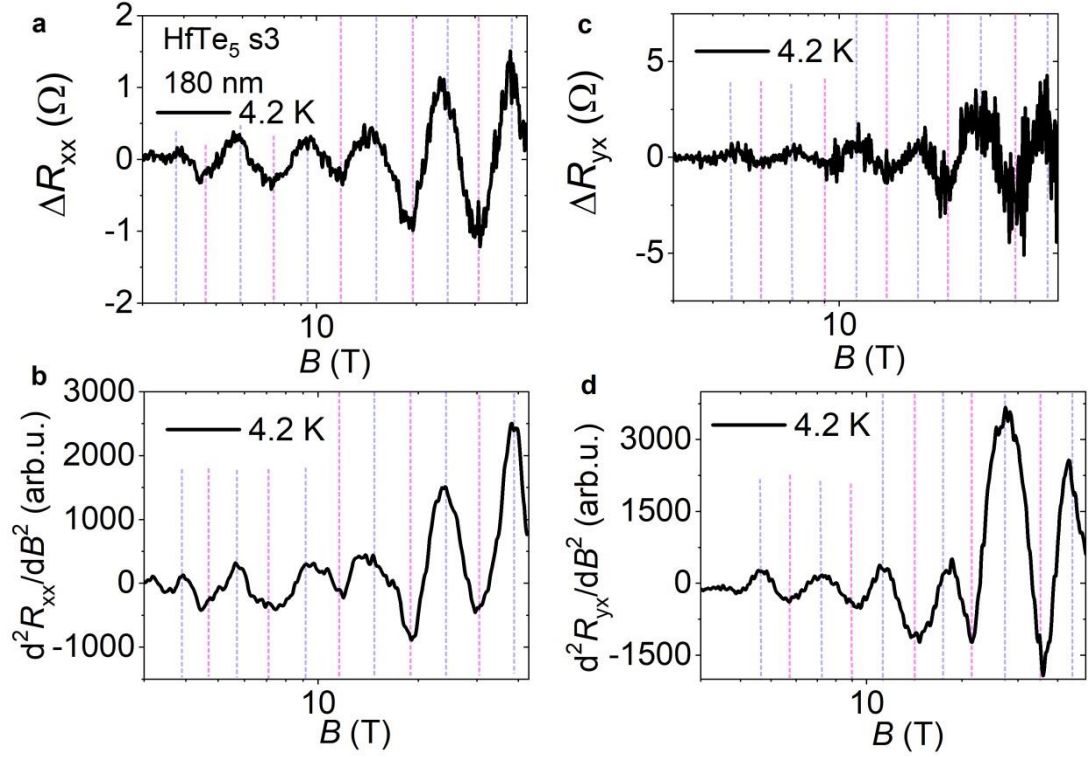

Supplementary Figure 3. Extracted log$B$-periodic oscillations in HfTe$_5$ flake (s3) at 4.2 K. (a) Log$B$-periodic oscillations in MR of s3 at 4.2 K from the raw data after subtracting a smooth background. (b) The second derivative results of the raw MR data of s3 at 4.2 K. (c) Log$B$-periodic oscillations in Hall traces of s3 at 4.2 K after subtracting a smooth background. (d) The second derivative results of the raw Hall trace of s3 at 4.2 K. Dashed lines serve as guides to the eye.

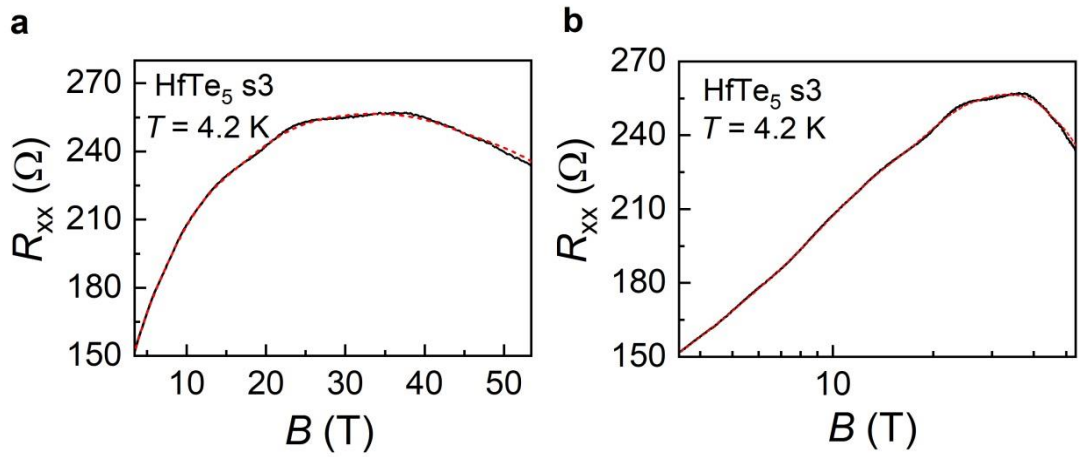

Supplementary Figure 4. Background obtained by smoothing the raw MR data of HfTe$_5$ s3 in a linear (a) and a semilogarithmic scale (b). The background and the raw MR data are shown by the red dashed line and the black points, respectively.

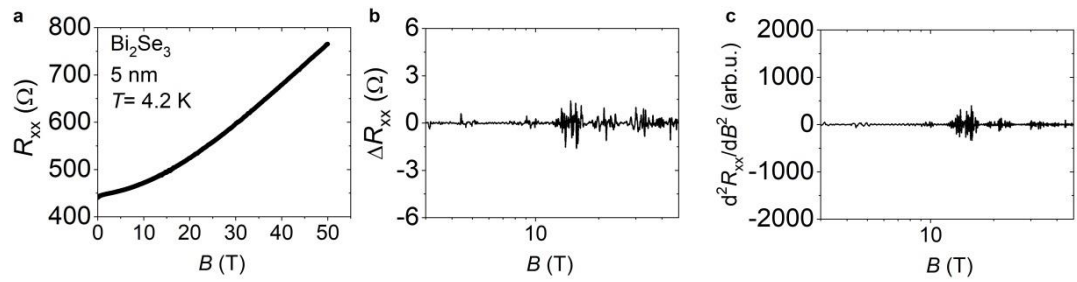

Supplementary Figure 5. Transport measurements of a $Bi_2Se_3$ film with a thickness of 5 nm in the same facility as $ZrTe_5$ and $HfTe_5$ flakes in the pulsed high magnetic field center. (a) MR of the $Bi_2Se_3$ film at 4.2 K. (b) Subtracting a smooth background of the data in (a). (c) Second derivative results of the data in (a). No oscillating feature is detected in the entire magnetic field region up to 50 T.

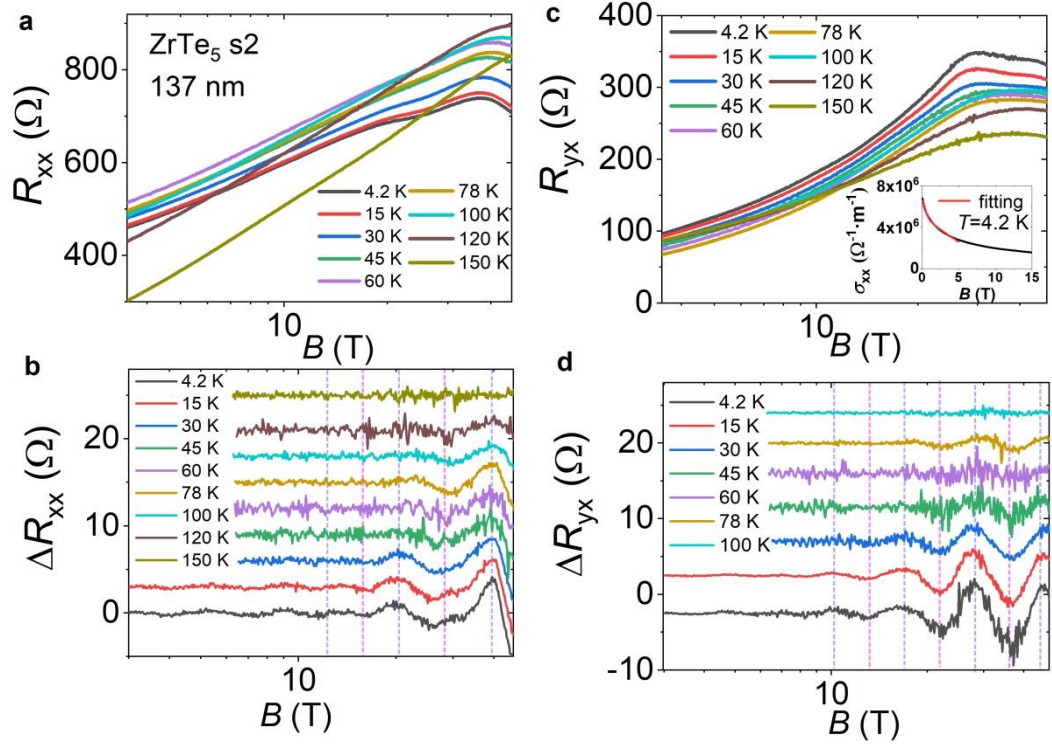

Supplementary Figure 6. Temperature-dependent log$B$-periodic oscillations in ZrTe$_5$ flake (s2) with a thickness of about 137 nm. (a) MR of s2 at different temperatures. (b) Log$B$-periodic oscillations in s2 at different temperatures. (c) and (d) Log$B$-periodic oscillations in Hall traces of s2 at various temperatures. Inset of (c): Two-carrier model fitting of s2 at 4.2 K. Red line is the fitting result. The Dirac hole density (mobility) is estimated to be $9.9 \times 10^{16}$ cm$^{-3}$ ($1.4 \times 10^{4}$ cm$^2$V$^{-1}$s$^{-1}$). Dashed lines serve as guides to the eye and data curves in (a) - (d) are shifted for clarity.

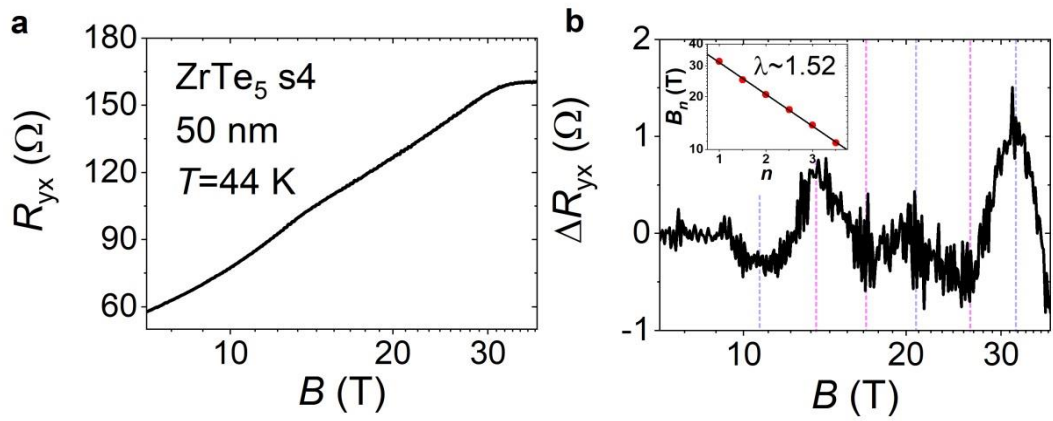

Supplementary Figure 7. Log-periodic oscillations and DSI in the ZrTe$_5$ flake (s4) with a thickness of about 50 nm. (a) Hall trace of s4 at 44 K. (b) Extracted oscillatory part from Hall trace of s4. The inset shows the log$B$-periodicity of the oscillations in s4. The scale factor $\lambda$ is about 1.52.

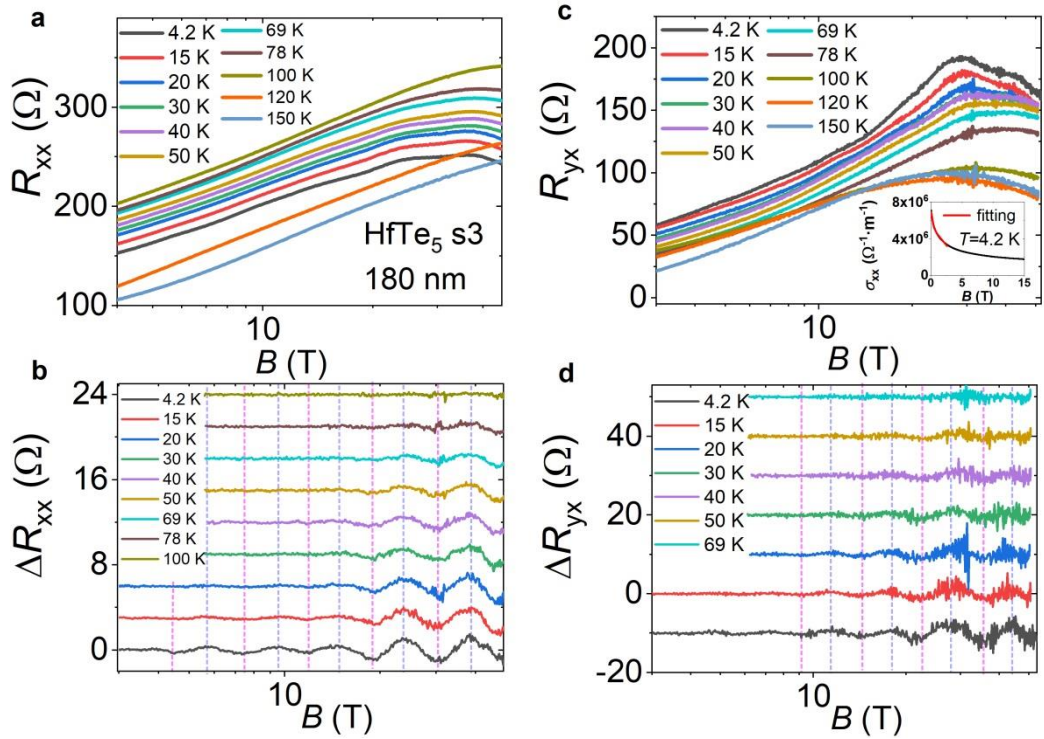

Supplementary Figure 8. Temperature-dependent log$B$-periodic oscillations in HfTe$_5$ flake (s3) with a thickness of about 180 nm. (a) MR of s3 at different temperatures. (b) Log$B$-periodic oscillations in s3 at different temperatures. (c) and (d) Log$B$-periodic oscillations in Hall traces of s3 at various temperatures. Inset of (c): Two-carrier model fitting of s3 at 4.2 K. Red line is the fitting result. The Dirac hole density (mobility) is estimated to be $5.0 \times 10^{16}$ cm$^{-3}$ ($3.0 \times 10^4$ cm$^2$V$^{-1}$s$^{-1}$). Dashed lines serve as guides to the eye and data curves in (a) - (d) are shifted for clarity.

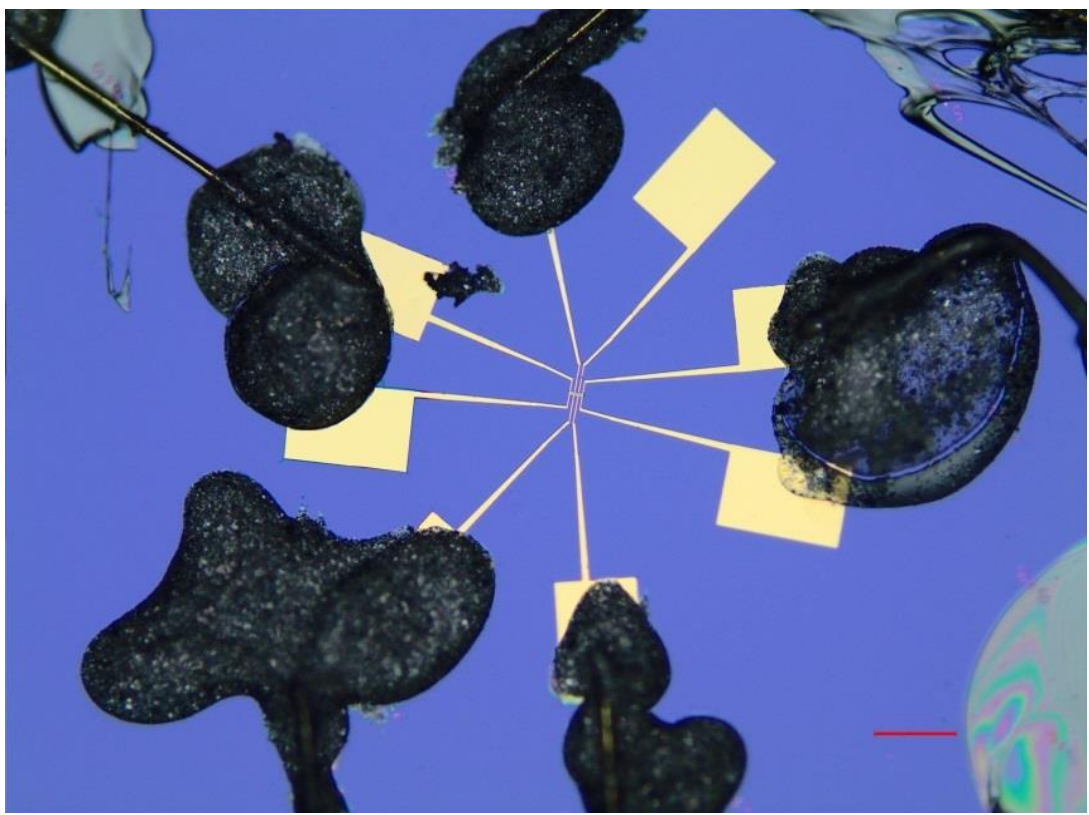

Supplementary Figure 9. Optical image of one typical ZrTe$_5$ flake device. Gold wires are attached to the electrodes by the silver epoxy and fixed on the substrate by GE varnish. Scale bar represents 100 μm.

## Supplementary References